\documentclass[journal=mamobx,manuscript=article]{achemso}
\usepackage[T1]{fontenc}
\usepackage[latin9]{inputenc}
\usepackage{textcomp}
\usepackage{amsthm}
\usepackage{amsmath}
\usepackage{graphicx}
\usepackage{amstext}
\usepackage{bm}

\newcommand{\bq}{\begin{equation}}
\newcommand{\eq}{\end{equation}}

\author{Danijel~Grgi\v{c}in}
\email{danijelgrgicin@gmail.com}
\phone{+385 98 9523 914}

\author{Damir~Vurnek}

\title{Universal conductivity dependence of pure water polyelectrolyte solutions}

\abbreviations{DNA, DS, SAXS}
\keywords{Conductivity, polyelectrolytes, mesh size, electric potential, dielectric spectroscopy}

\begin{document}
	\begin{figure*}
		\includegraphics[width=83mm]{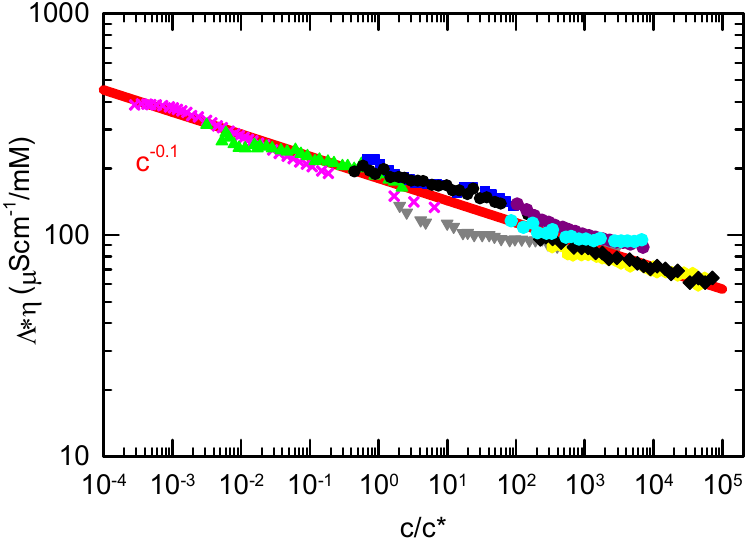}
		\caption*{Table of Contents}
		\nonumber
		\label{Table of Contents}
	\end{figure*}
	
	\begin{abstract}
		In order to understand how live matter functions one needs to understand the interaction between polyelectrolytes. We discover a general dependence of polyelectrolyte conductivity valid in at least nine decades of polyelectrolyte concentration spanning dilute and semidilute pure water solutions. Furthermore, we showed that current state of the art theories can not explain polyelectrolyte conductivity and suggest the path in transport theories which needs to be taken in order to explain polyelectrolyte conductivity.
	\end{abstract}
	
	\section*{Introduction} 
	Polyelectrolytes are extremely complicated systems due to myriad mutually competing effects. This makes them hard to study but they are a basis of life which makes them also extremely important. Study of polyelectrolyte conductivity begins in mid 20's century. Researchers noticed much smaller conductivity as compared to conductivity of equal amount of electrolyte\cite{huizenga}. Later that was explained as due to competing of electrostatic energy of charges imprisoned on polyion and  entropy of counterions in solution leading to accumulation of counterions in polyion vicinity in order to screen polyion excessive electrostatic energy. This effect was first understood by Manning\cite{manning69,oosawaknjiga71,manning78,manningconductivity}. But even when accounted for this effect theories could not explain conductivity data\cite{wandrey1999,wandrey2000}. 
	
	This phenomenological theory assumes that total electric conductivity $\sigma$ of aqueous polyelectrolyte solution is given as a sum of conductivity contribution of all charged species:
	\begin{equation}
	\sigma=\sum_i|z_i|c_i\lambda_i
	\label{conductivity1}
	\end{equation}
	\noindent
	where $c_i$ is molar concentration of the charge specie of type $i$, valency $|z_i|$ and molar conductivity $\lambda_i$\cite{colbyreview,cametti}. For simple pure water polyelectrolyte consisting of charged polyions (p) and counterions (c) Eqn. \ref{conductivity1} can be reduced to
	
	\begin{equation}
	\sigma=c(\lambda_p^c + \lambda_c^c)
	\label{equivalentconductivity}
	\end{equation}
	
	where $c$ is monomer concentration which is in the case of DNA equal to phosphate concentration and $\lambda_p^c$ is monomer conductivity. 
	
	\noindent
	For strongly charged polyelectrolytes counterion condensation occurs reducing conductivity $\eta^{-1}$ times. 
	\noindent
	\begin{equation}
	\Lambda \equiv \sigma/c=\eta^{-1}(\lambda_p^c + \lambda_c^c)
	\label{equivalentconductivity2}
	\end{equation}
	
	The charge-density parameter $\eta$ needs  more attention. Manning-Oosawa (MO)\cite{manning69,oosawaknjiga71,manning78,manningconductivity} modeled the polyelectrolyte as a rigid rod with linear charge separation $b$ immersed in solution with relative permittivity $\varepsilon_r$. Bjerumm length $l_B$ is the distance at which electrostatic energy of two charges with valencies $z_1$ and $z_2$ equals their entropic energy $l_B=\frac{z_1z_2e^2}{4\pi \varepsilon_0 \varepsilon_ r\ k_BT}$. It depends on solution temperature $T$ and for two monovalent charges immersed in water at 25$^{\circ}$C it has the value of $l_B=0.72\ nm$. Together with linear charge separation $(b)$ it defines the charge density parameter $\eta=l_B/b$ which determines the onset of MO condensation. If $\eta>1$ counterions condense on the polyion until they effectively reduce the linear charge separation to $l_B$, i.e until $\eta$ becomes equal to 1. In table \ref{tab1} we summarized the parameters needed to determine $\eta$ for polymers we used in experiments/data analysis.
	
		\begin{table}
			\centering
			\caption{$b$, $l_b$ and $\eta$ for polyelectrolytes used in this study.}
			\begin{tabular}{cccc}
				\hline 
				&$b(e^-/nm)$&$l_b (nm)$ & $\eta$  \\
				\hline
				Na-DNA & 0.17 & 0.72 & 4.24 \\
				ss Na-DNA & 0.43 & 0.72 & 1.68 \\
				Mg-DNA & 0.17 & 1.44 & 8.4\\		
				Na-HA & 1 & 0.72& 1 \\
				Mg-HA & 1 & 1.44& 1.44 \\
				Na-PSS & 0.25 & 0.72& 2.88 \\
				PDADMAC& 0.53& 0.72&1.35 \\
				SO$_4^{-2}$ PDADMAC &0.53 & 1.44&2.7 \\
				poly(vinylbenzyltri- & 0.25& 0.72 &2.88 \\
				alkylammonium) chloride & & & \\
				\hline				
			\end{tabular}
			\label{tab1}%
		\end{table}
		
	Still equation \ref{equivalentconductivity2} did not fully fix the conductivity problem. Next effect for which was thought that would fix this mismatch was the conductivity contribution of polyion conformation\cite{colbyreview,cametti}. Cametti used Dobrynin's description of polyion conformation\cite{dobrynin} to include its contribution to total conductivity \cite{cametti}. But even with this upgrade experimental results cannot be explained with such state-of-the-art conductivity theories\cite{colbyreview,cametti}. Beyond this effect numerous more effects are incorporated into conductivity equation, which now has more than 15 different symbols, like eqn 83 from Ref. \citenum{colbyreview}, but still experimental data can not be explained\cite{wandrey2000}. Later in this report we will show that conductivity of both dilute and semidilute polyelectrolytes can be modeled with a single formula for broad range of conditions. We obtained it on polydisperse (long) DNA, polydisperse poly(styrene sulphonate) PSS, $146bp$ DNA and polydisperse hyaluronic acid (HA) in $0.02mM<c<50mM$ monomer concentration range with sodium and magnesium counterions. 
	
	In the past non fitting results greatly improved our understanding of polyelectrolytes, i.e. smaller than expected conductivity of polyelectrolytes lead to the realization that counterions condense on the polyion. We hope our current measurements can push the field in the same direction.
	
	\section*{Materials and Methods}
\begin{figure}
		\includegraphics[width=89.0mm]{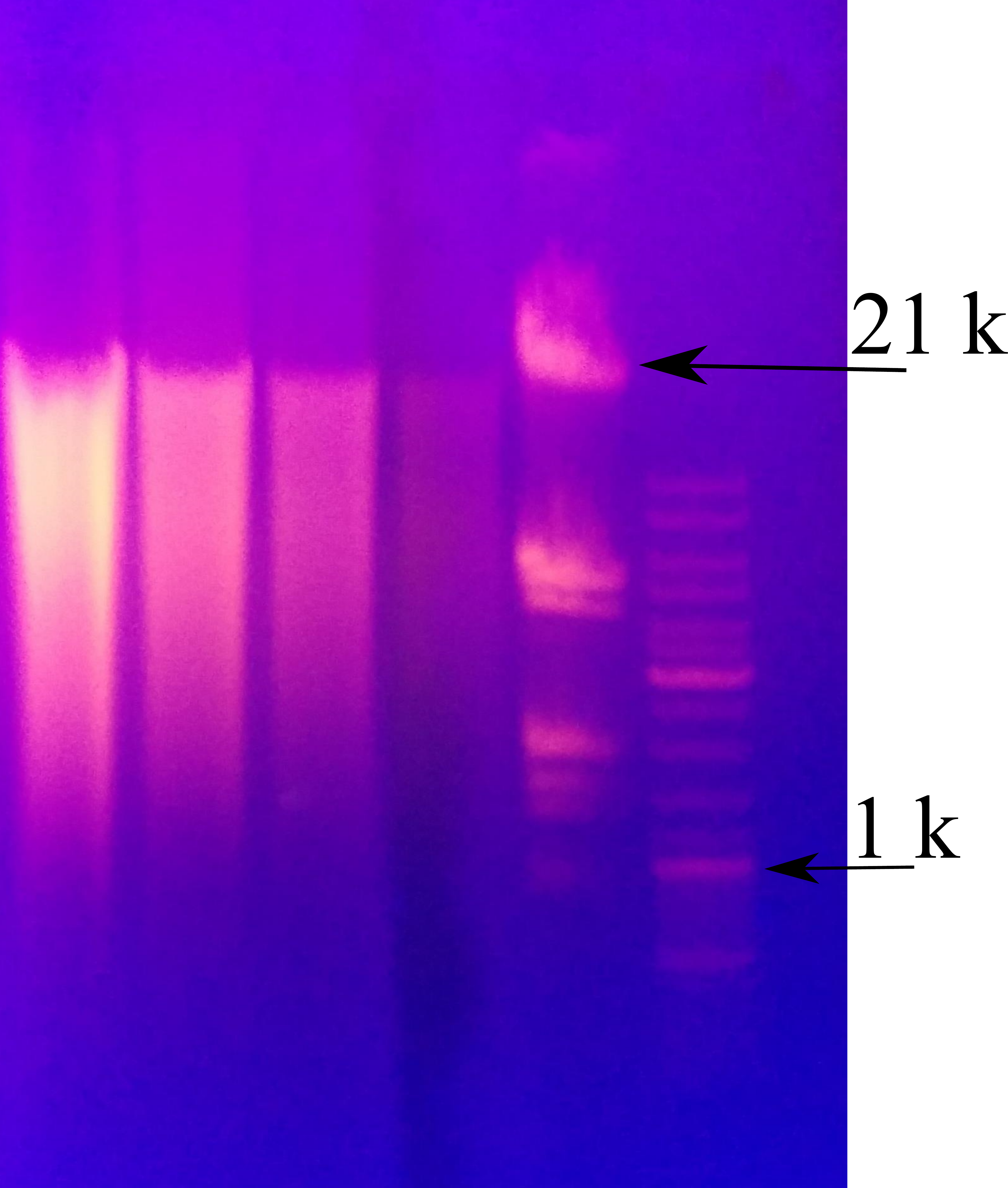}
		\caption{The representative image of agarose gel with DNA samples in the first four lanes and ladders in lines 5 and 6. DNA quantities in lines are 5, 2, 1 and 0.5 $\mu g$ and ladders are thermo scientific SM0191 and thermo scientific SM0333 respectively.}
		\label{fig5}
	\end{figure}

	146bp nucleosomal DNA was obtained as described in our previous publication\cite{vuletic11}, while Hyaluronic acid (53747), polydisperse (long) DNA (D1626) and long PDADMAC (409030) was obtained from Sigma-Aldrich. Na-PSS($Mp=679000$, $Mw=666000$) was obtained from PSS. We have measured conductivity of pure water samples with Na$^+$ or Mg$^{2+}$ counterions and compared it with corresponding samples with added salt (sodium or magnesium). Pure water sodium samples were produced by dissolving obtained threads or powders in Milipore water ($\sigma<1.5\mu S/cm$) following protocol I from Ref.\citenum{pre07}. To produce magnesium/SO$_4^{-2}$ samples additional step was needed in which sodium/clor ions were removed by multiple dialysis against MgCl$_2$/H$_2$SO$_4$ of higher concentration. Solutions with added salt were produced following protocol II3 from Ref. \citenum{pre07}, except for long Mg-DNA for which additional dialysis was introduced to produce samples with even lower ionic strength of added salt $I_s=0.033\ mM$. Ionic strength of produced samples can be found in Appendix A. Monomeric concentration of produced samples was in range $0.02\ mM<c<50\ mM$. Polydispersity of obtained lyophilized DNA treads was in the range of 1-20 kbp, as can be seen on Fig. \ref{fig5}. During production of pure water Na-DNA samples special caution needs to be taken in order not to deal with partially denatured samples on which more information can be found in Appendix B.  
	
	Concentration of dilute semidilute crossover $c^*$ depends on monomer size $b$ and number of monomers in polymer $N$\cite{salamon,combet}
	\begin{equation}
	c^*=\frac{m_m}{A b^3 N^2}
	\label{crossover}
	\end{equation}
	
	where $m_m$ is molar mass, $A$ is Avogadro number. For shorter polyions $c^*$ has larger value. Determination of $c^*$ for polydisperse samples, like our polydisperse DNA and PSS is not straight forward. In order to calculate crossover concentration we take $N=1kbp$ and $N=3296$ for DNA and PSS respectively. Producer of DNA did not specify polymer size and $N=1kbp$ was taken as lowest polymer size obtained on gel electrophoresis. In the case of PSS producer specify average number molar mass $M_p= 679 000$ which was taken for calculation of crossover concentration. For PDADMAC producer specified average number molar mass as  $M_p= 400 000 - 500 000$. Crossover concentrations were 4mM, 0.0001mM, 0.021mM, 0.0098 mM, 0.0008 mM for 146bp DNA, HA, long polydisperse DNA, PSS and PDADMAC respectively. 
	For conductivity measurement we use homemade conductivity chamber whose distance between parallel platinum plates is $l=0.1021\pm 0.0001cm$ and chamber constant corresponding to sample volume of $100\ \mu L$ is $l/S=0.1042\pm 0.008cm^{-1}$. We use conductivity data obtained at 100 kHz with Agilent 4294A impedance analyzer. 100 kHz is chosen because at such high frequency there is no influence of electrode polarization which plagues data at lower frequencies. Ac amplitude of $50\ mV$ is chosen to probe samples after establishing that measurements are independent of applied voltage in the range of 5-1000 mV. More detailed description of the technique can be found in our earlier publications\cite{pre07,vuletic10,physicab}. All measurements are done at 25.00 $\pm\ 0.01^{\circ}$C.

	\section*{Results}
	\begin{figure}
		\includegraphics[width=89.0mm]{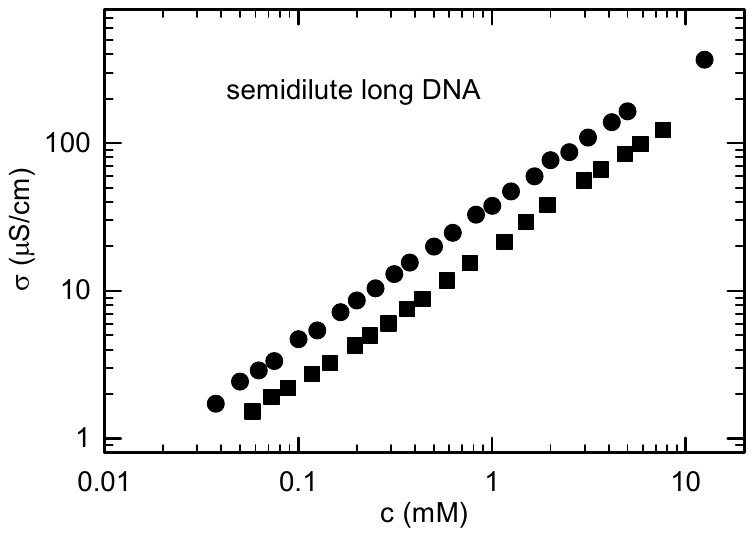}
		\caption{Conductivity of pure water solutions of Na-DNA (circles) and Mg-DNA (squares) as a function of nucleotide concentration.}
		\label{fig1}
	\end{figure}
	
\begin{figure}
		\includegraphics[width=89.0mm]{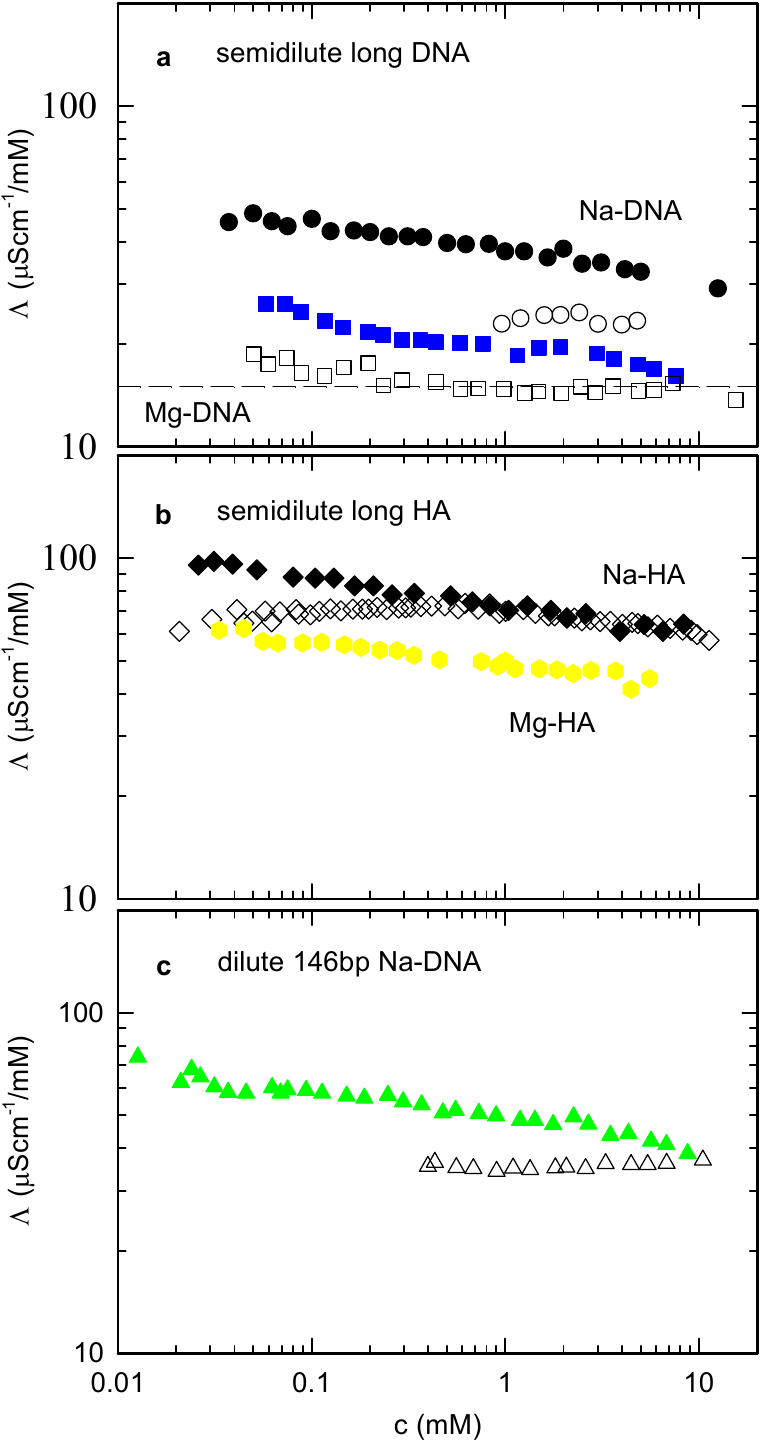}
		\caption{{\bf a}, Conductance of Na-DNA (black circles) and Mg-DNA (blue squares) deduced from pure water (full) and added salt (open symbols) solutions. Dashed line is horizontal guide for the eye. {\bf b}, Conductance of Na-HA (black diamonds) and Mg-HA (yellow hex) in pure water (full) and deduced from solution with 0.1 mM added NaCl (open symbols). {\bf c}, Conductance of 146bp Na-DNA (full green triangle up) in pure water and deduced from solution added 1 mM NaCl (open symbols).}
		\label{fig2}
	\end{figure}

	Conductivity $\sigma$ of pure water semidilute Na-DNA and Mg-DNA solutions as a function of nucleotide (phosphate) concentration $c$ is shown in Fig. \ref{fig1}. As predicted by Eqn. \ref{conductivity1} conductivity increases with the increase of polyion concentration. Conductivity of strongly charged polyelectrolytes with divalent counterions is smaller than that of monovalent as predicted by Eqn. \ref{equivalentconductivity2}. In chosen concentration range conductivity changes more than 100 fold so in order to precisely compare the values of all measured polyelectrolytes on Fig. \ref{fig2} we show conductance ($\Lambda$) vs concentration ($c$) for various polyelectrolytes.
	
	Conductance of pure water Na-DNA and Mg-DNA samples decreases with increasing concentration while the corresponding conductance deduced from solution with added salt does not. According to Egn. \ref{conductivity1} all charged species independently contribute to conductivity. That means if we subtract added salt contribution from conductivity we should get exactly the value of conductivity of pure water polyelectrolyte. But on Fig. \ref{fig2} we see that conductivities obtained with these two procedures differ. That directly tells us that Eqn. \ref{conductivity1} is not correct. Furthermore, since decrease of conductivity with increasing concentration is not correctly theoretically predicted researchers usually wrongly ascribe that decrease to concentration dependence of $\eta^{-1}$ in order to preserve "validity" of Eqn. \ref{conductivity1} \cite{das2012, das2016,nyquist,netz03,netz13}. If the assumption that $\eta^{-1}(c)$ is correct we should observe it in pure water as well as in solutions with added salt. But our conductivities of samples with added salt are not concentration dependent. Moreover, the same decrease we observe also in samples for which condensation does not occur (Na-HA). Therefore we conclude that the cause of conductivity dependence on concentration is not in the concentration dependence of $\eta^{-1}$, i.e. $\eta^{-1}$ is not a function of concentration at least in the concentration range which we measured.  
	
	\section*{Discussion}
\begin{figure}
		\includegraphics[width=89.0mm]{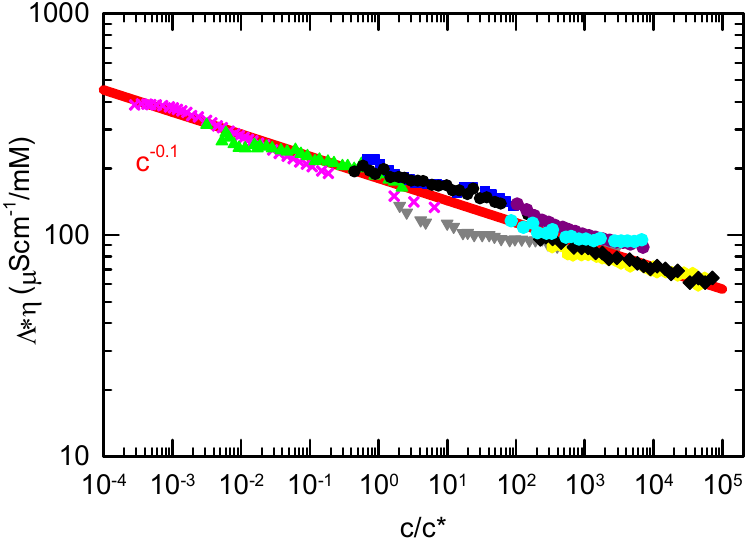}
		\caption{Master curve of renormalized conductivity as a function of renormalized concentration. Symbols represent the same polyelectrolytes as on Fig. \ref{fig2} with addition of Na-PSS sample (dark grey triangle down), PDADMAC (dark pink circle), SO$_4^{-2}$ PDADMAC (cyan circle) and data for poly(vinylbenzyltrialkylammonium) chloride (pink x) which we have taken from Ref. \citenum{wandrey2000}. Red line is $\propto (c/c^*)^{-0.1}$.}
		\label{fig3}
	\end{figure}

	With a suitable transformation we noticed that all measured conductivities lie on the same master curve, seen Fig. \ref{fig3}.
	
	\noindent
	We would like to point out that data on the master curve span nine decades and happen on both sides of crossover concentration. The most deviation from master curve can be observed for $1<c/c^*<100$, i.e. in the semidilute regime right after passing the crossover concentration. We ascribe this deviation to polydispersity of our samples. All samples that we measure in that range (Na-PSS, long Na and Mg-DNA and PDADMAC) are polydisperse. Main problems with a polydisperse sample is determination of the $c^*$. One can expect that such problem would have the biggest impact in vicinity of crossover concentration, where indeed our data most deviate from the master curve, and smaller impact as one moves away from the crossover concentration, where our data do not deviate from the master curve.
	Based on Fig. \ref{fig3} we propose a simple conductivity dependence that would be valid universally for pure water solutions 
	
	\begin{equation}
	\Lambda=C\eta^{-1} (c/c^*)^{-0.1}.
	\label{model}
	\end{equation}
	\noindent
	
	Here we emphasize that we achieve this result on dilute (146bp Na and Mg-DNA) and semidilute (long Na and Mg-DNA, Na and Mg-HA and Na-PSS) solutions, with weakly (Na-HA) and strongly (146bp Na-DNA, long Na and Mg-DNA, Mg-HA and Na-PSS) charged polyions and on different types of polyions withe either negative (DNA, HA and PSS) or positive (PDADMAC) charges on them. Although one can expect that $C$ is a function of a lot of variables it is interesting that for all polyelectrolytes shown in Fig. \ref{fig3} it has the same value. Furthermore, combining informations on Fig \ref{fig2}, Fig \ref{fig3} and the Eqn. \ref{model} we concluded the following:
	
	1) Models based on Eqn. \ref{conductivity1} can not explain our data. That is especially evident in discrepancy of conductivity for pure water samples and samples with added salt provided on Fig \ref{fig2}.

	2) According to Egn. \ref{equivalentconductivity} counterions as well as polyions contribute to conductivity. But according to our data on Fig. \ref{fig3} all different polyions, with different shapes bearing negative as well as positive charges should have the same $\lambda_p^c$. That brings the question about mechanism which could comply all such properties. Positive and negative charged polyions on Fig. \ref{fig3} lay on the same line which means that same mechanism governs conductivity. Positively charged polyion could lead electrons but it is hard to imagine how negatively charged polyions lead positrons. If instead they lead positive ions that should be much slower and conductivity should be much lower as compared to electron leading conductivity suggesting that polyion charges does not contribute to total conductivity. Evidence pointing in the same direction can be found by comparing our experiments done with monovalent and divalent counterions. The same polyion with divalent counterions have 2 times smaller conductivity as compared to polyion with monovalent counterions. But according to Eqn \ref{equivalentconductivity2} in both that cases $\lambda_p^c$ is equal. It is hard to rationalise mechanism of polyion conductivity in which $\lambda_p^c$ does not changes when linear charge of the equal polyion changes two times. 
	Thirdly, if there is some longitudinal conductivity component most probable it would not be linear with longitudinal linear charge density for molecules which radius is more than order of magnitude greater than linear charge density. In the case of DNA molecule diameter is 2nm and linear charge density is 1e-/0.17 nm or in the case of divalent counterions 1e-/0.34. That is almost order of magnitude smaller than radial distance of these charges since they are situated on outer diameter of DNA. From all these arguments we conclude that $\lambda_p^c=0$.     

	3) It seems like the only counterion parameter that enters the equation is counterion valence, trough $\eta$. But it has been known for a long time that polyelectrolyte conductivity depends also on the counterion radii \cite{jan,janet}, so most probably had we have chosen counterions with more different radii we would have observed that $C$ depends also on counterion radii.   

	4) When renormalized by charge density parameter $\eta$ conductivity of both strongly and weakly charged polyions lie on the master curve. This suggests that the conductivity mechanism we observe is the same for all samples and could be explained with a same theory/equation for both strongly and weakly charged polyions. 

\begin{figure}
		\includegraphics[width=89.0mm]{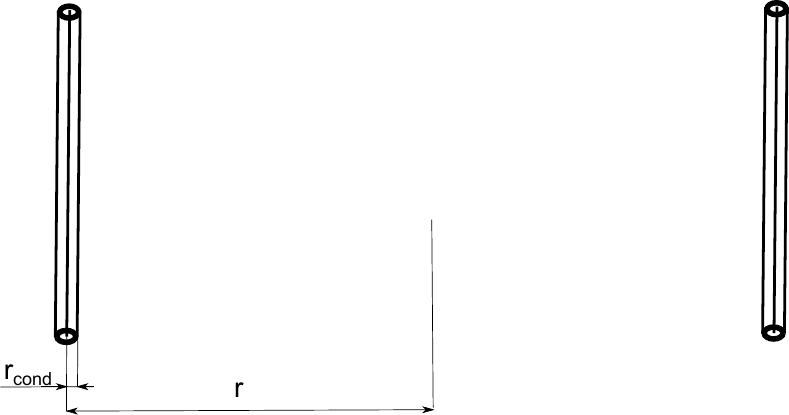}
		\caption{Polyelectrolyte composed of two different regions as regarding counterion diffusion. In the smaller region $(V_{cond})$ are condensed counterions which do not participate in conductivity while in the rest of the space $V_{pol}(r)$ are diffusive counterions which diffusion coefficient are governed with distance from the polyion so they are not completely free.}
		\label{figmodel}
	\end{figure}

	5) The nature of concentration dependence $(c/c^*)^{-0.1}$ stays unclear. Typically counterions are divided into two groups: condense and "free". It is assumed that condensed counterions do not participate in osmotic pressure and conductivity, and that "free" counterions are all identical and that each of them equally participate in osmotic pressure/conductivity. Many conductivity data, including data in this report, both in dilute and semidilute solutions can not be successfully explained with such assumptions. Alternative possibility is that non condensed counterions are not all identical and that their contribution to conductivity is dependent on their distance from the polyion. In this picture counterions further away from polyion would be more free and their contribution to conductivity would increase. Here is a simple 2D analogy with people in busy shopping center. Imagine almost full shopping center, like before Christmas holidays. People without shopping carts represent water molecules while people with shopping charts represents ions with neutralising water layers. When they are in the vicinity of some pillar (representing polyion) they slow down in order to pass behind pillar, naturally people with shopping charts slow down more. By increase number of peoples with shopping charts(counterions) and number of pillars (polyions) people with shopping charts slow down more frequently. Now we are back to solution. The average counterion-polyion distance decreases with increasing concentration, as we illustrate this on Fig. \ref{figmodel} and we argue that this directly leads to decreasing $\Lambda$ 
	
	As regarding counterion diffusion polyelectrolyte is divided in two regions. Condensed counterions are situated in polyion vicinity. They do not contribute to conductivity, i.e. their diffusion coefficient is zero $D_{cond}=0$. The volume they take we call $V_{cond}$. In the remaining space diffusion coefficient rise as one get further away from the polyion $D_{pol}(r)>D_{cond}$. With regard to that we suggest the foolowing model:

	\begin{equation}
	\Lambda\propto \eta^{-1} D_{pol}(c)
	\label{model1}
	\nonumber
	\end{equation}
	\noindent

	\section*{Conclusion}
	State-of-the-art theories cannot explain our conductivity data. All these phenomenological theories assume that total conductivity can be taken as sum over all charged species, and that their contributions are independent of each other. We showed that this assumption is not valid.
	Further we show our newly discovered empirical law between normalized conductance and normalized polyion concentration $\Lambda \eta \propto (c/c^*)^{-0.1}$ valid in nine decades of concentration on both sides of crossover concentration. We have showed that this concentration dependence cannot be due to $\eta$ since it should be concentration independent. We have also showed that $\lambda_p^c$ is identical for all polyions we tested. 
	
	The prevailing simple view used in effective conductivity theories is that the polyelectrolyte solutions are composed of fully condensed and fully free counterions. Our data suggest that a different model is more suitable. In salt free solutions part of the counterions closer to the polyion significantly feels the influence of the polyion, i.e. counterions are not fully free and the degree of their freedom depends on the distance from the polyion. Counterions further away from the polyion are more free.

	\section{Appendix A}
	In order not to confuse with polyion concentration we express added salt as ionic strenghts  $I_S=1/2 \sum_i c_i z_i^2$. Ionic strength and corresponding Debye screening length $\kappa^{-1}(nm)\approx 10/\sqrt{I_s(mM)}$ of the solutions used in this report can be found in Tab. \ref{tab2}
	
	\begin{table}[h!]
		\centering
		\caption{Concentration (c(mM)), ionic strength $(I_s(mM))$ and corresponding Debye screening length $(\kappa^{-1})$ of solutions used in this report.}
		\begin{tabular}{cccc}
			\hline 
			sample & $c(mM)$ & $I_S(mM)$ & $\kappa^{-1}(nm)$ \\
			\hline
			Na-DNA & 1 & 1 & 10\\
			Mg-DNA & 0.033 & 0.1 &  30\\
			Na-HA & 0.1 &   0.1 & 30\\		
			146bp Na-DNA & 1 & 1 & 10\\
			146bp Mg-DNA & 1 & 3 & 5.5\\
			\hline				
		\end{tabular}
		\label{tab2}%
	\end{table}

	\section{Appendix B}
	\begin{figure}
		\includegraphics[width=82.5mm]{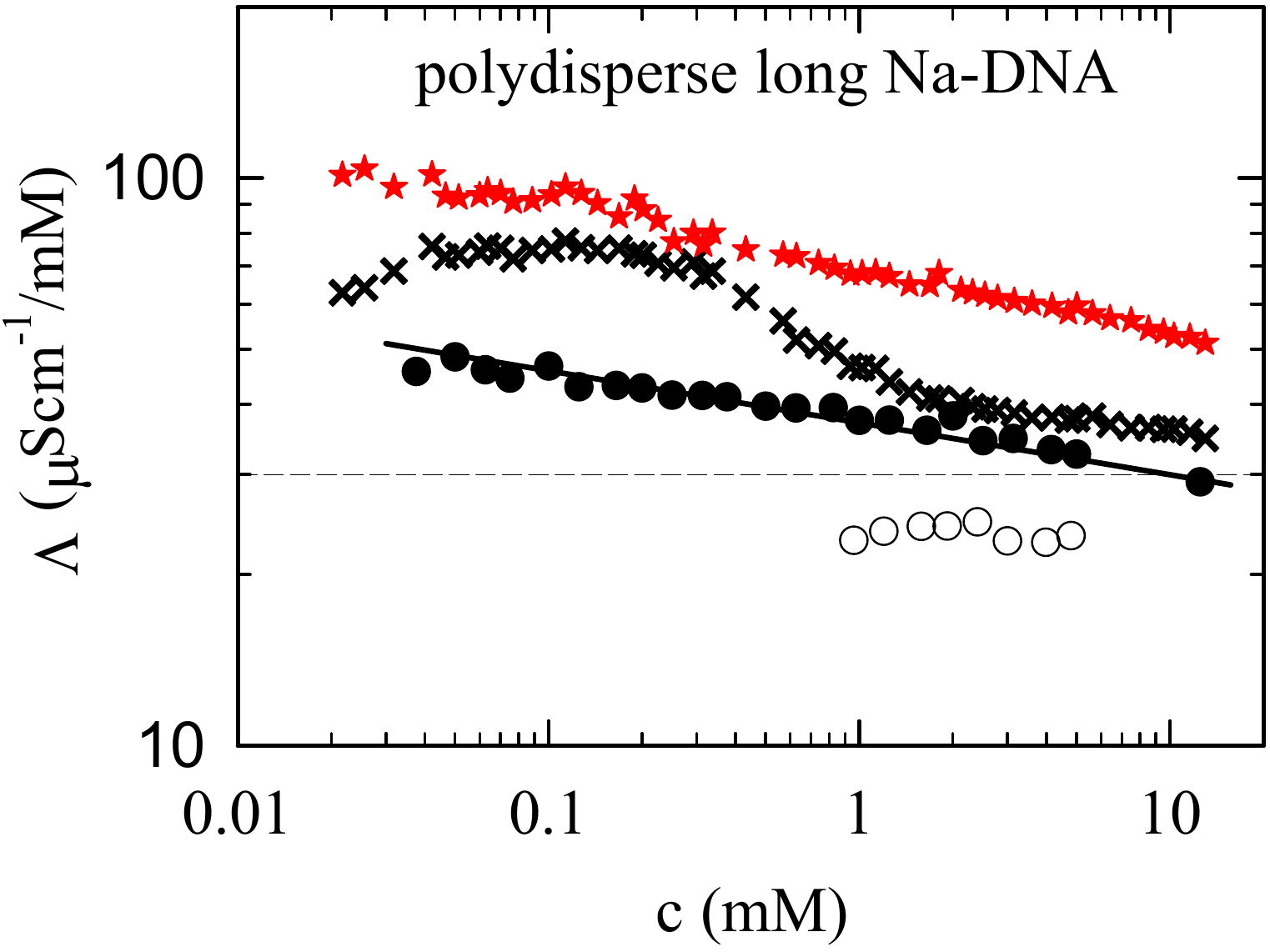}
		\caption{Conductance of long NaDNA samples as function of concentration. Symbols are as in Fig. \ref{fig3} with addition of denatured samples (stars) and native samples which spontaneously partially denature(cross).}
		\label{fig7}
	\end{figure}
	
	UV spectrophotometry enables determination of conformation of DNA in solution since ssDNA absorbs around 40\% more electromagnetic radiation at 260 nm than dsDNA\cite{bloomfield}. More about UV spectrophotometry can be found in our previous publication\cite{physicab}. With UV spectrophotometry we check conformation of DNA and only for Na-DNA pure water samples we occasionally noticed denaturation. Such samples are not included in this report since denaturation influence conductance as can be seen on Fig. \ref{fig7}. In order to determine span of conductivity range which denatured samples might have we perform full denaturation by heating samples to 97°C. This data is shown in figure \ref{fig7} with stars.

\end{document}